\documentclass[runningheads]{llncs}
\usepackage{amsfonts}
\usepackage{graphicx}
\usepackage[pagebackref=true,breaklinks=true,letterpaper=true,colorlinks,bookmarks=false]{hyperref}

\begin{document}
\title{Symmetric-Constrained Irregular Structure Inpainting for Brain MRI Registration with Tumor Pathology}
\titlerunning{Symmetric Irregular Inpainting for Brain Tumor MRI Registration}
\author{Xiaofeng Liu\inst{1\dag} \and Fangxu Xing\inst{1\dag} \and Chao Yang\inst{2} \and C.-C. Jay Kuo\inst{3} \and Georges El Fakhri\inst{1} \and Jonghye Woo\inst{1}}

\institute{Gordon Center for Medical Imaging, Department of Radiology, Massachusetts General Hospital and Harvard Medical School, Boston, MA, 02114, USA\and
Facebook Artificial Intelligence, Boston, MA, 02142\and
Ming Hsieh Department of Electrical and Computer Engineering, University of Southern California, Los Angeles, CA, 90007, USA\\$\dag$ Contribute Equally.}

\authorrunning{X. Liu et al.}

\maketitle              % typeset the header of the contribution

\vspace{-20pt}\begin{abstract}
Deformable registration of magnetic resonance images between patients with brain tumors and healthy subjects has been an important tool to specify tumor geometry through location alignment and facilitate pathological analysis. Since tumor region does not match with any ordinary brain tissue, it has been difficult to deformably register a patient’s brain to a normal one. Many patient images are associated with irregularly distributed lesions, resulting in further distortion of normal tissue structures and complicating registration's similarity measure. In this work, we follow a multi-step context-aware image inpainting framework to generate synthetic tissue intensities in the tumor region. The coarse image-to-image translation is applied to make a rough inference of the missing parts. Then, a feature-level patch-match refinement module is applied to refine the details by modeling the semantic relevance between patch-wise features. A symmetry constraint reflecting a large degree of anatomical symmetry in the brain is further proposed to achieve better structure understanding. Deformable registration is applied between inpainted patient images and normal brains, and the resulting deformation field is eventually used to deform original patient data for the final alignment. The method was applied to the Multimodal Brain Tumor Segmentation (BraTS) 2018 challenge database and compared against three existing inpainting methods. The proposed method yielded results with increased peak signal-to-noise ratio, structural similarity index, inception score, and reduced L1 error, leading to successful patient-to-normal brain image registration.\vspace{-5pt}
\end{abstract}

\section{Introduction}\vspace{-5pt}

In brain imaging studies, magnetic resonance imaging (MRI) as a noninvasive tool is widely used to provide information on the brain’s clinical structure, tissue anatomy, and functional behaviors~\cite{oishi2010mri,bauer2013survey}. When multiple datasets from a population of interest are involved, to establish a comparable framework in which similarity and variability in the tissue structure can be evaluated, deformable image registration between subjects are often used to achieve inter-subject alignment~\cite{sotiras2013deformable}. Brain tumor is a common type of disorder diagnosed using medical imaging~\cite{sartor1999mr}. However, tumors in MRI tend to cause difficulties with deformable registration: 1) Tumor regions have no matching structure in a normal brain, nullifying the basic mathematical assumptions made for regular image registration methods and subsiding their performance; 2) Expansion of tumor regions often alters its peripheral structure, causing the whole image to become asymmetric with distorted hemispheres or ventricles; and 3) The locations of tumors are sometimes irregularly scattered around the whole brain, causing inconsistencies when matching multiple tumor spots~\cite{deangelis2001brain}.

%\begin{figure}[t]
%\begin{center}
%\includegraphics[width=1\linewidth]{fig//miccai1.pdf}
%\end{center}\vspace{-15pt}
%\caption{Brain tumor inpainting results of a few MRI slices of a patient.}
%\label{fig:1}\vspace{-10pt}
%\end{figure}

There has been a great deal of work that tackles patient-to-normal tissue registration in a traditional way \cite{tang2017groupwise,lamecker:inria-00616156}. Especially, for small tumor cases, Dawant et al.~\cite{dawant2002brain} introduced a tumor seed and Cuadra et al.~\cite{cuadra2002atlas} extended it with a tumor growth model to drive the registration process. For larger tumors, Mohamed et al.~\cite{mohamed2006deformable} used a biomechanical model of tumor-induced deformation to generate a similar tumor image from the normal image. Since then many methods have been focusing on tumor growth simulations to facilitate symmetry computation~\cite{gooya2010deformable,zacharaki2008orbit}. More traditional methods are summarized in~\cite{sotiras2013deformable}. In this work, we propose a new image inpainting method---i.e., a restorative method that treats tumor as defective holes in an ideal image and reconstructs them with synthetic normal tissue. The synthesized brain can be processed with regular deformable registration and the tumor region will eventually be re-applied after being mapped to the new space.

Traditional inpainting methods are either diffusion-based or patch-based with low-level features~\cite{barnes2009patchmatch,ballester2001filling,bertalmio2000image,criminisi2004region,efros2001image,prados2016fully}. These prior approaches usually perform poorly in generating semantically meaningful contents and filling in large missing regions~\cite{liu2019coherent}. Recently developed learning-based inpainting methods usually use generative adversarial networks (GANs) to learn image semantics and infer contents in the missing region~\cite{iizuzuka2017globally,song2018contextual,pathak2016context,yang2018image}. In the brain tumor application, difficulties 2) and 3) need to be addressed specifically. Starting from the initial context encoder deep learning method~\cite{pathak2016context}, Liu et al.~\cite{liu2018image} updated the mask and convolution weights in each layer to handle irregular holes. However, it is challenging for these 1-step inpainting solutions to address the large holes with complicated texture. Song et al.~\cite{song2018contextual} proposed a multi-step framework to refine the results with patch-swap, but its coarse inpainting module does not fit for multiple irregular holes. Moreover, the above methods are designed for general image cases and do not involve priors such as brain anatomy and physiology.

In this work, we propose a novel multi-step inpainting method capable of making fine-grained prediction within irregular holes with feature patch-wise conditional refinement. It also incorporates a symmetry constraint to explicitly exploit the quasi-symmetry property of the human brain for better structure understanding. Deformable registration is applied between inpainted patient images and normal controls whose deformation field is then used to deform original patient data into the target space, achieving patient-to-normal registration.

% a novel learning-based inpainting method that adopts image-to-image translation~\cite{isola2016image} as a coarse inpainting stage. It is 

\section{Methods}\vspace{-5pt}

Given a brain MRI slice $I_0$ with tumor, the goal is to replace the pathological regions with normal brain appearances. The incomplete input $I_0$ is composed of  $R$ and $\overline{R}$, representing the removed pathological region (the hole) and the remaining normal region (boundary or context), respectively. Mathematically, the task is to generate a new, complete image $I$ with plausible contents in $\overline{R}$. 

\begin{figure}[t]
\begin{center}
\includegraphics[width=1\linewidth]{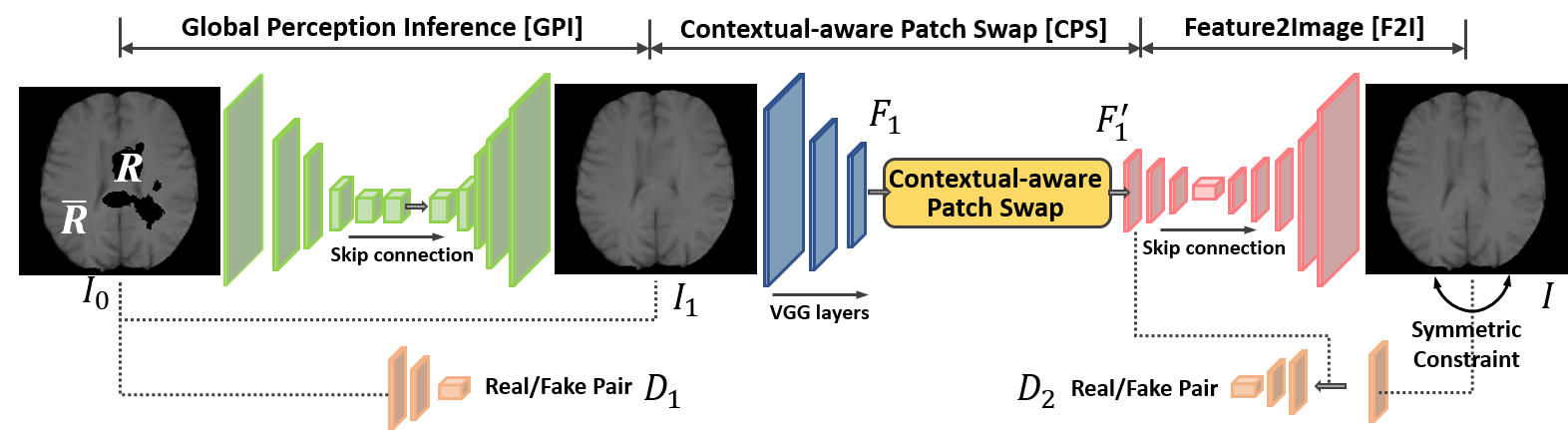}
\end{center}\vspace{-15pt}
\caption{Overview of the proposed network architecture. GPI is used for coarse inference and VGG is used for extracting the feature map. The patch-swap layer propagates high frequency information from the boundary to the hole. F2I translates to a complete, high-resolution image further constrained with symmetric loss.}
\label{fig:2}\vspace{-5pt}
\end{figure}

Following the basic idea of contextual-based image inpainting~\cite{song2018contextual}, our framework consists of three sequential modules: global perception inference (GPI), context-aware patch swapping (CPS), and feature-to-image translator (F2I). The intuition behind the multi-step operation is that direct learning of the distribution of high dimensional image data is challenging. Thus using a coarse generation followed by a refinement scheme can increase the inpainting performance~\cite{song2018contextual}. Our network architecture is shown in Fig.~\ref{fig:2}.

\subsection{Global Perception Inference}

The input to the GPI network $I_{0}$ is a 1$\times$240$\times$240 image with irregular holes. Its output is a coarse prediction $I_1$. Considering the potential irregular distribution of tumor locations, the rectangular hole generation module used in \cite{song2018contextual} is not applicable. Therefore, we first adopt the GPI network structure from the image-to-image translation network proposed in \cite{isola2017image}, which consists of 4$\times$4 convolutions with skip connections in order to concatenate different features from each encoder layer and the corresponding decoder layer. We slightly modify the size of each layer since only single channel T1-weighted MRI is used in this task.

% To Xiaofeng: I_1, I_{gt} not defined. And why is I_{gt} written in parentheses (I_{gt}) in eqn.(1)?

The GPI module is explicitly trained using the $L_1$ reconstruction loss, which is important for stabilizing the adversarial training \cite{liu2018normalized}. It can be formulated as \begin{eqnarray}
\mathcal{L}_{1}(I_1, I_{gt}) &=& \parallel I_1-I_{gt}\parallel_1,\vspace{-5pt}
\end{eqnarray}
\label{eqn:l1}where $I_1$ and $I_{gt}$ are the rough inpainting result of GPI and the ground truth, respectively.

The second objective is the adversarial loss based on GANs \cite{liu2019feature}, which can be defined as:\vspace{-5pt}\begin{eqnarray}
\mathcal{L}_{adv} = \max_{D_1}\mathbb{E}[\log(D_1(I_0, I_{gt}))+\log(1-D_1(I_0, I_1))].\vspace{-5pt}
\end{eqnarray}
Here, a pair of images are input to the discriminator $D_1$ as is the setting of adversarial training.  The incomplete image $I_0$ and the original image $I_{gt}$ are the real pair, and the incomplete image $I_0$ and the prediction $I_1$ are the fake pair.

During training, the overall loss function is given by $
\mathcal{L}_{GPI} = \lambda_{1}\mathcal{L}_{1}+\lambda_{2}\mathcal{L}_{adv}$, where $\lambda_1$ and $\lambda_1$ are the balancing hyperparameters for the two losses. 

\subsection{Context-aware Patch Swapping}

We use $I_{1}$ as input to the CPS network which is implemented in two phases. First, $I_{1}$ is encoded as $F_1$ by a fully convolutional network (FCN) using the pre-trained VGG network as in \cite{song2018contextual}. Then the patch-swap operation is applied to propagate the texture from $\overline{R}$ to $R$ while maintaining the high frequency information in $R$ \cite{liu2019permutation}. 

\begin{figure}[t]
\begin{center}\vspace{-5pt}
\includegraphics[width=0.8\linewidth]{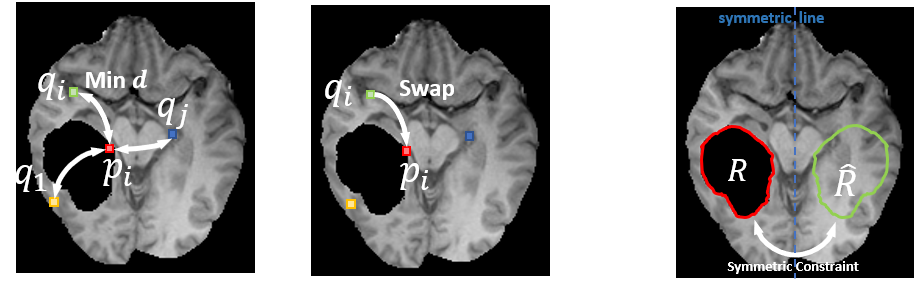}
\end{center}\vspace{-15pt}
\caption{Illustration of the patch-swap operation (left) and symmetry constraint (right). Patch-swap is implemented in the FCN-based VGG's feature space to search for the most similar boundary $1\times1$ feature patch with minimum $d(p,q)$.}
\label{fig:3}\vspace{-15pt}
\end{figure}

$r$ and $\bar{r}$ denote the regions in $F_1$ corresponding to $R$ and $\bar{R}$ in $I_1$, respectively. For each 1$\times$1 neural patch\footnote{In the inpainting community, the 1$\times$1 patch (in a feature map) is a widely used concept. The output of F1 $\in\mathbb{R}^{256\times60\times60}$, while the original image is 240$\times$240$\times$1; therefore a 1$\times$1 area in a feature map is not considered as a pixel.} $p_i$ of $F_1$ overlapping with $r$, the closest-matching neural patch in $\bar{r}$, indexed by $q_i$, is found using the following cross-correlation metric\vspace{-5pt}
\begin{eqnarray}
d(p,q) =\frac{<p,q>}{\parallel p\parallel\cdot \parallel q\parallel}\vspace{-5pt},
\end{eqnarray} where $p_i$ is replaced by $q_i$. We first swap each patch in $r$ with its most similar patch in $\bar{r}$, followed by averaging overlapping patches. The output is then a new feature map $F'_1$. This process is illustrated in Fig.~\ref{fig:3} left.

%Each neural patch in the hole $r$ searches for the most similar neural patch on the boundary $\bar{r}$, and then swaps with that patch.

\subsection{Feature-to-image Translator}

Next, we use the F2I network to learn the mapping from the swapped feature map to a complete and vivid image, which has a U-Net style generator. The input to the U-Net is a feature map extracted by the FCN-based VGG network. The generator consists of seven convolution layers and eight deconvolution layers, where the first six corresponding deconvolutional and convolutional layers are connected through skip connections. The output is a complete 1$\times$240$\times$240 image. In addition, the F2I network comprises a patch-GAN based discriminator $D_2$ for adversarial training. However, the input to $D_2$ is a pair of an image and its feature map in contrast to the GPI network.

In practice, we follow \cite{song2018contextual} that uses the ground truth as training input. Specifically, the feature map $F_{gt}=\mbox{vgg}(I_{gt})$ is the input to the patch-swap layer followed by using the swapped feature $F'_{gt}=\mbox{patch\_swap}(F_{gt})$ to train the F2I model. $F'_1=\mbox{patch\_swap}(F_1)$ is still used as input for inference, since $I_{gt}$ is not accessible at test time. Of note, using different types of input for both training and testing is not a common practice in training a machine learning model. However, its effectiveness in inpainting has been demonstrated in \cite{song2018contextual}. Similar to \cite{zheng2016improving}, the robustness can be further improved by sampling from both the ground truth and the GPI prediction. 

The first objective is the perceptual loss defined on the entire image between the final output $I$ and the ground truth $I_{gt}$:\begin{eqnarray}
\mathcal{L}_{perceptual}(I, I_{gt}) =\parallel vgg(I)-vgg(I_{gt})\parallel_2. 
\end{eqnarray} This perceptual loss has been widely used in many tasks \cite{gatys2016image,johnson2016perceptual,dosovitskiy2016generating,chen2016fast} as it corresponds better with human perception of similarity~\cite{zhang2018unreasonable}. 

The adversarial loss is defined by the discriminator $D_2$, which can be expressed as: \vspace{-5pt}
\begin{eqnarray}
\mathcal{L}_{adv} = \max_{D_2}\mathbb{E}[\log(D_2(F'_{gt}, I_{gt}))+\log(1-D_2(F'_{gt}, I))],
\end{eqnarray} where the real and fake pairs for adversarial training are ($F'_{gt},I_{gt}$) and ($F'_{gt},I$), respectively.

\subsection{Quasi-symmetry Constraint}

While the brain is not exactly symmetrical w.r.t. the mid-sagittal plane, there is a large degree of symmetry between left and right hemispheres in the brain which we call the ``quasi-symmetry property" \cite{raina2019exploiting,oostenveld2003brain}. As such, using this anatomical symmetry constraint on the generated images can mitigate the ill-posed inpainting task and further improve performance especially for large hole cases. The symmetry loss is given by \begin{eqnarray}
\mathcal{L}_{sym}(I)  = \mathbb{E} \parallel   I_{R }- I_{\hat{R}}   \parallel_2 ,
\end{eqnarray}where $R$ and $\hat{R}$ are the hole and its mirrored region as shown in Fig.~\ref{fig:3} right. 

Therefore, we can easily transfer the appearance of the normal brain tissue to the corresponding tumor part by teaching the network to recover the lost information from the mirrored side. Note that the brains used in our experiments are coarsely aligned on their mid-sagittal planes. More importantly, our technique is robust against any potential misalignments, since the resolution of the feature space is 60$\times$60, while the input is 240$\times$240, with the down-sampling of the maxpooling operation, the deep neural networks, in general, are robust against small rotation \cite{marcos2016learning}. Besides, the deep neural network can tackle this simple rotation. With the symmetry constraint, the overall loss for the F2I translation network is defined as:\vspace{-5pt}\begin{eqnarray}
\mathcal{L}_{F2I} = \lambda_{3}\mathcal{L}_{perceptual}+\lambda_{4}\mathcal{L}_{adv}+\lambda_{5}\mathcal{L}_{sym},
\end{eqnarray}where $\lambda_3$, $\lambda_4$, and $\lambda_5$ are the balancing hyperparameters for different losses. Considering the brain is not strictly symmetrical w.r.t. the mid-sagittal plane, we usually choose a relatively small weight $\lambda_5$ for $\mathcal{L}_{sym}$.

\section{Experiments and Results}
\begin{figure}[t]
\begin{center}
\includegraphics[width=1\linewidth]{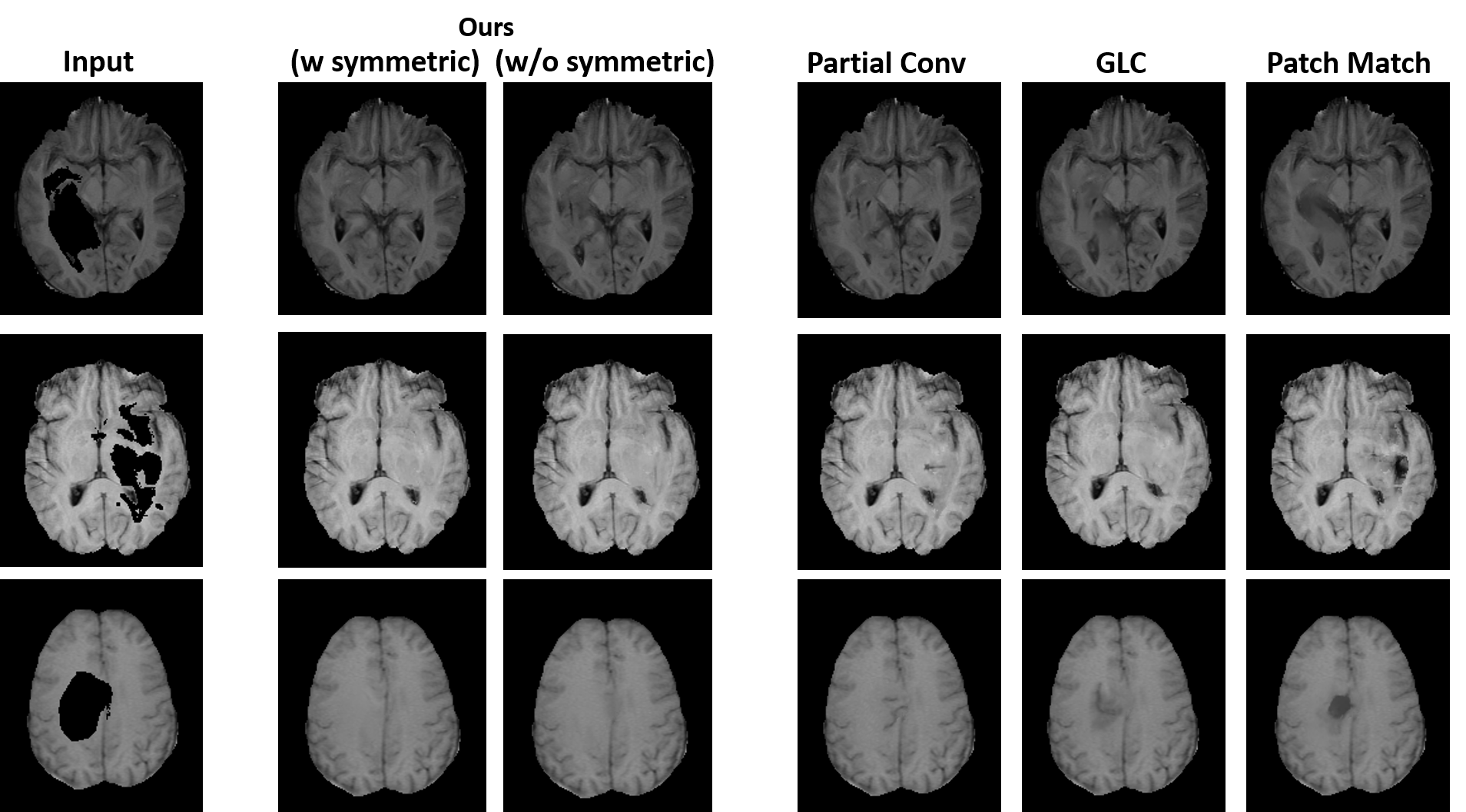}
\end{center}\vspace{-20pt}
\caption{An ablation study of our symmetry constraint and the comparison with the other inpainting methods.}\vspace{-10pt}
\label{fig:5}
\end{figure} 

The proposed method was validated both qualitatively and quantitatively on the T1 modality of Brain Tumor Segmentation (BraTS) 2018 database\footnote{\url{https://www.med.upenn.edu/sbia/brats2018/data.html}}. From a total of 210 patients each with \~{}150 slices, we randomly selected 16 patients for testing and the remaining subjects were used for training in a subject independent manner. Training was performed on four NVIDIA TITAN Xp GPUs with the PyTorch deep learning toolbox~\cite{paszke2017automatic}, which took about 5 hours.

The normal slices without tumors in the training set were selected to train our network. Since tumors in BraTS data can occur in different spatial locations, our network is capable of familiarizing with the normal appearance in different slices. We randomly chose the irregular tumor segmentation labels in our training set as training masks.

The process of computing cross-correlation for all the neural patch pairs between the hole and the remaining region (e.g., boundary) is computationally prohibitive. To alleviate this, the strategy in~\cite{chen2016fast,song2018contextual} was used to speed up computation via paralleled convolution. In practice, processing one feature map only took about 0.1 seconds.

In order to match the absolute value of each loss, we set different weights for each part. For the training of GPI, we set weight $\lambda_{1}=10$ and $\lambda_{2}=1$. Adam optimizer was used for training. The learning rate was set at $lr_{GPI}=1\mathrm{e}{-3}$ and $lr_{D_1}=1\mathrm{e}{-4}$ and the momentum was set at 0.5. When training the F2I network, we set $\lambda_{3}=10$, $\lambda_{4}=3$ and $\lambda_{5}=1$. For the learning rate, we set $lr_{F2I}=2\mathrm{e}{-4}$ and $lr_{D_2}=2\mathrm{e}{-4}$. Same as the GPI module, the momentum was set as 0.5.

\begin{figure}[t]
\begin{center}
\includegraphics[width=1\linewidth]{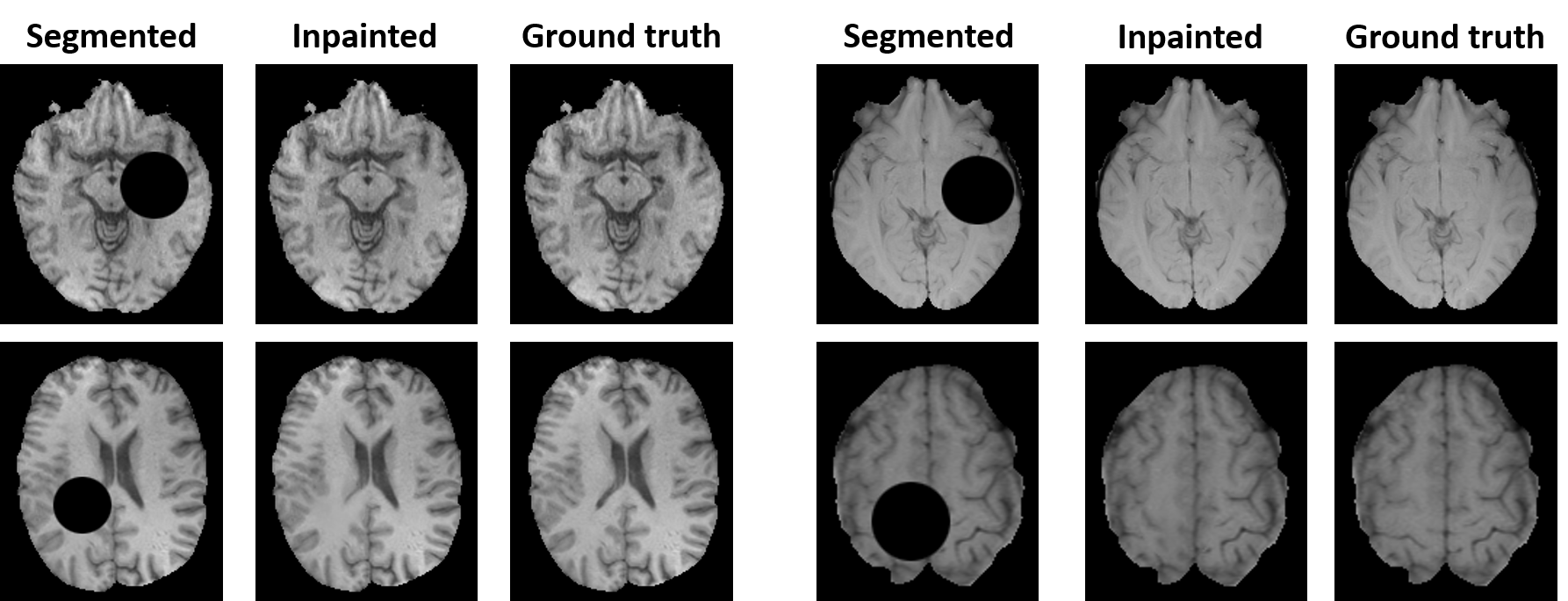}
\end{center}\vspace{-15pt}
\caption{Inpainting results comparing to the ground truth.}
\label{fig:6}
\end{figure}

\begin{figure}[h]
\begin{center}
\includegraphics[width=1\linewidth]{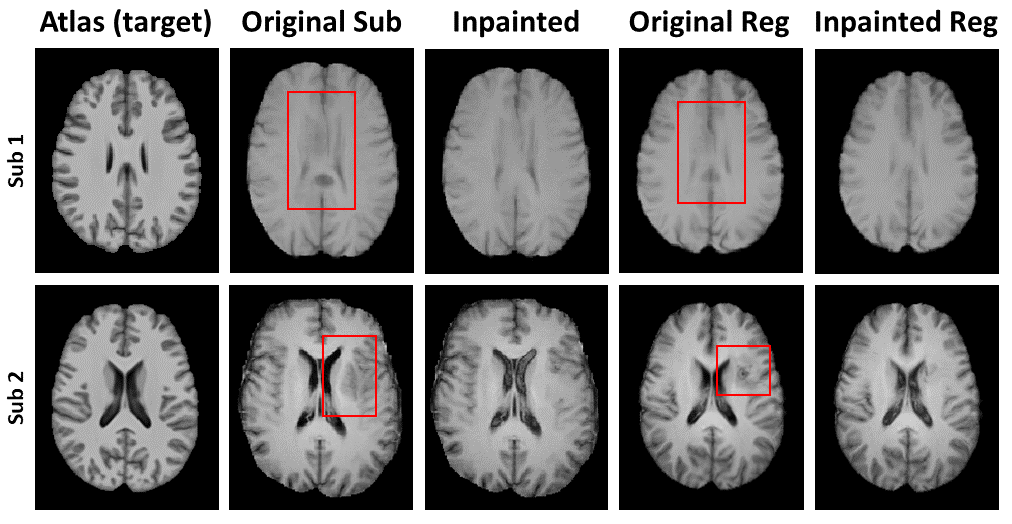}
\end{center}\vspace{-15pt}
\caption{Deformable registration of two brain tumor subjects to a brain atlas: direct registration vs. inpainted registration. Tumors are marked in red boxes.} 
\label{fig:7}
\end{figure}

The inpainting results of various cases are shown in Figs.~\ref{fig:5}, \ref{fig:6}, and \ref{fig:7}. The proposed network can deal with incomplete data from different unseen patients, different slice positions, and arbitrary shape and number of holes.

Comparisons with the other inpainting methods are shown in Fig.~\ref{fig:5}. Our proposed method using context-aware inpainting \cite{song2018contextual} shows superior performance over the other methods as visually assessed. In addition, an ablation study to evaluate the contribution of the symmetry constraint is illustrated in Fig.~\ref{fig:5}. Of note, the inpainting quality was further improved using the symmetry constraint plus marginal training cost without additional testing cost. This is partly attributed to the use of the context and quasi-symmetry property of the brain. 

\begin{table}[t]
\caption{Numerical comparison of four methods using BraTS 2018 testing set. Note that smaller mean L1 error and larger SSIM mean error indicate higher similarity.}\label{tabel:1}\vspace{-5pt}
\centering
\begin{tabular}{c|c|c|c|cccc|ccc}
\hline
Methods& ~mean L1 error~$\downarrow$ ~~&~~~SSIM~$\uparrow$~~&	~~~PSNR~$\uparrow$~~	& Inception Score $\uparrow$\\\hline\hline
Patch-match \cite{barnes2009patchmatch}	&445.8&	0.9460&	29.55&	9.13\\
GLC \cite{iizuka2017globally}	&432.6&	0.9506&	30.34&	9.68\\
Partial Conv \cite{liu2018image}	&373.2&	0.9512&	33.57&	9.77\\\hline
Proposed	&292.5&	0.9667&	34.26&	10.26\\
Proposed+symmetry&	\textbf{254.8}&	\textbf{0.9682}	&\textbf{34.52}&	\textbf{10.58}\\
\hline
\end{tabular}
\end{table}

\begin{table}[h]
\caption{Mutual information between registered brain volumes and the brain atlas on ten test subjects using direct patient registration and inpainted volume registration.}\vspace{-5pt}
\centering
\resizebox{1\linewidth}{!}{%
\begin{tabular}{c|c|c|c|c|c|c|c|c|c|c}
\hline
Methods& Sub1 & Sub2 & Sub3 & Sub4 & Sub5 & Sub6 & Sub7 & Sub8 & Sub9 & Sub10 \\
\hline\hline
Direct registration	&0.303&0.311&0.308&0.324&0.315&0.299&0.309&0.303&0.317&0.308\\
Inpainted registration &0.309&0.311&0.309&0.324&0.316&0.304&0.312&0.313&0.320&0.312\\
\hline
\end{tabular}%
}
\vspace{-10pt}
\label{tabel:2}
\end{table} 

For quantitative evaluation, we manually generated holes with random size and positions on normal slices of the testing subjects. Therefore, the ground truth is known. The inpainted images were expected to have sharp and realistic looking textures, be coherent with $\bar{R}$, and look similar to its corresponding ground truth. Our results are illustrated in Fig.~\ref{fig:6}. The proposed method generated visually satisfying results. Table~\ref{tabel:1} lists numerical comparisons between the proposed approach, Patch-match \cite{barnes2009patchmatch}, GLC \cite{iizuka2017globally}, and Partial Conv \cite{liu2018image}. We note that the compared inpainting baselines \cite{iizuka2017globally,liu2018image} are based on the 1-step framework. We used four quality measurements to assess the performance: mean L1 error, structural similarity index (SSIM), peak signal-to-noise ratio (PSNR), and inception score~\cite{salimans2016improved}. We directly computed the mean L1 error and SSIM over the holes, while the incepetion score is measured on the completed $I$.

Finally, Fig.~\ref{fig:7} and Table~\ref{tabel:2} show the results of deformable registration using the ANTs SyN method~\cite{avants2011reproducible} with normalized cross-correlation as a similarity metric. As for the target atlas, we used a T1-weighted brain atlas constructed using healthy subjects from the OASIS database~\cite{marcus2007open}. The result was evaluated using mutual information (MI) computed only in normal tissues to achieve a fair comparison (tumor masks were used to exclude the tumor region). Direct patient-to-normal registration was affected by the existence of tumor, thus reducing the MI score even in normal tissues. This was corrected by using the inpainted volume as registration input, yielding improved or equal MI scores on every subject tested. The mean of MI was improved from 0.3097 to 0.3129.

\section{Conclusion}

This paper presented an inpainting network that replaces the pathological tumor regions with normal brain appearances, targeting patient-to-normal deformable registration. The challenges lie in irregular brain tumor distribution. The two-stage inpainting scheme utilized both the complete and segmented samples, producing the refined results based on pixel-wise semantic relevance. Our experimental results demonstrate that the proposed method surpassed the comparison methods, which can be used for the registration between healthy subjects and tumor patients.

\section{Acknowledgements}

This work was supported by NIH R01DE027989, R01DC018511, R01AG061445, and P41EB022544.

\bibliographystyle{splncs04}
\bibliography{egbib}

\end{document}